\documentclass[%
 reprint,
 amsmath,amssymb,
 aps,
 pra, 
 longbibliography
]{revtex4-2}

\usepackage{footmisc,multirow}
\usepackage{subfigure} 
\usepackage{amsmath,slashed}
\usepackage{upgreek}
\usepackage{graphicx}
\usepackage{comment}

\begin{document}

\title{
Addendum: Multichannel quantum defect theory of strontium bound Rydberg states (2014 J. Phys. B: At. Mol. Opt. Phys. 47 155001)}

\author{C. L. Vaillant}
\affiliation{Department of Physics, Durham University, South Road, Durham DH1 3LE, United Kingdom}
\author{M. P. A. Jones}
\affiliation{Department of Physics, Durham University, South Road, Durham DH1 3LE, United Kingdom}
\author{R. M. Potvliege}
\email{r.m.potvliege@durham.ac.uk}
\affiliation{Department of Physics, Durham University, South Road, Durham DH1 3LE, United Kingdom}


\begin{abstract}
Newly calculated multichannel quantum defect theory parameters and channel fractions are presented for the singlet and triplet S, P and D series and singlet F series of strontium. These results correct those reported in Vaillant C L, Jones M P A and Potvliege R M 2014 {\it J.
Phys. B: At. Mol. Opt. Phys.} {\bf 47} 155001.
\end{abstract}

\maketitle

It has recently been drawn to our attention that some of the results we reported in Ref.~\cite{Vaillant2014} are not reproducible with the MQDT parameters stated in that article. The issue could be traced to an erroneous sign in a program used for several of the series considered, which made these parameters incorrect.
This error does not affect the analysis of the $5\mbox{s}n\mbox{s}\,^1\mbox{S}_0$ and $5\mbox{s}n\mbox{d}\,^{1,3}\mbox{D}_2$ states outlined in \cite{Vaillant2014}. It does not invalidate the discussion of the radiative lifetime of these states given in that article either or the discussion of interchannel F\"orster resonances in the calculation of $C_6$ dispersion coefficients given in a subsequent publication \cite{Vaillant2015}. However, the MQDT parameters quoted in \cite{Vaillant2014} are incorrect for many of the series, as are some of the conclusions on the importance of particular channels in certain series.  

The primary aim of the present communication is to correct these results where necessary. However, we are taking the opportunity of this revision for integrating recently published results in the set of the experimental spectroscopic data at the basis of our MQDT models. These new data come from high precision measurements of the $5\mbox{s}n\mbox{s}\,^3\mbox{S}_1$, $5\mbox{s}n\mbox{d}\,^{3}\mbox{D}_1$ and $5\mbox{s}n\mbox{d}\,^{3}\mbox{D}_2$ series and of a few states of the $5\mbox{s}n\mbox{d}\,^{1}\mbox{D}_2$ series \cite{Couturier2019,Tan2022}. The energies of singlet S, P, D and F states have also been measured recently, for principal quantum numbers between approximately 50 and 70 \cite{Brienza2023}; however, these states lie beyond the range of energies amenable to the present calculations.

We have revisited all the series for which MQDT models were reported in \cite{Vaillant2014}, namely all the singlet and triplet S, P and D series as well as the singlet F series. We have calculated new sets of MQDT parameters for all these series. With a few exceptions, described below, these new models are based on the same choice of channels and the same or similar sets of experimental data as in our original publication. The channels and experimental data considered for each series are listed in table~\ref{table:channels} and in table~\ref{table:data}, respectively.
We fitted the singlet and triplet D$_2$ states together in view of the strong mixing between these two series.
We used the values given in \cite{Couturier2019} for the first ionisation threshold and the mass-corrected Rydberg constant of $^{88}$Sr, i.e., $I_s = 45932.2002$~cm$^{-1}$ and $\tilde{R} = 109736.631$~cm$^{-1}$, which supersede the values used in \cite{Vaillant2014}.

Compared to \cite{Vaillant2014}, the main difference in the choice of experimental energies is our use of the results of \cite{Couturier2019}, in preference to older data, for some of the $5\mathrm{s}n\mathrm{d}\,^1\mathrm{D}_2$ states and for the $5\mathrm{s}n\mathrm{s}\,^3\mathrm{S}_1$, $5\mathrm{s}n\mathrm{d}\,^3\mathrm{D}_1$ and $5\mathrm{s}n\mathrm{d}\,^3\mathrm{D}_2$ states with $n \geq 12$. However, the experimental error of $3\times 10^{-5}$~cm$^{-1}$ on these results is two to four orders of magnitude smaller than for the other experimental energies included in the calculation, which caused difficulties with finding models fitting the data over the whole range of relevant states. In particular, most of the $5\mathrm{s}n\mathrm{d}\,^1\mathrm{D}_2$ experimental energies currently available are known only to within a much larger error of $1\times 10^{-3}$~cm$^{-1}$ \cite{Beigang1982a}. To avoid an imbalance between the singlet and triplet D$_2$ series, we increased the error on the results of \cite{Couturier2019} to $1\times 10^{-3}$~cm$^{-1}$, to match the error on most of the other energies included in calculation.
We did the same when fitting the $5\mathrm{s}n\mathrm{s}\,^3\mathrm{S}_1$ and $5\mathrm{s}n\mathrm{d}\,^3\mathrm{D}_1$ series, too, as is explained below.

The theoretical and numerical methods used in the present calculations are the same as in \cite{Vaillant2014}, with one minor difference, but the computer programs are newly developed and completely independent. The difference concerns the normalisation of the mixing coefficients $Z_i$: we now normalise these coefficients as per Eq.~(6.49) of \cite{Seaton1983}, for consistency with the energy-dependence of the $K_{ij}$ parameters, rather than as per Eq.~(21) of \cite{Vaillant2014}. (These mixing coefficients were denoted $\bar{A}_i$ in \cite{Vaillant2014}. They are denoted $Z_i$ here for conformity with the literature). The codes are published separately and are freely available \cite{PotvliegeCPC2023,mqdtfitGitHub}.

\begin{table}[htbp]
\caption{The dissociation channels included in the present MQDT models and their ionisation limit.}
\begin{ruledtabular}
\begin{tabular}{llr}
Series & \multicolumn{1}{l}{Channels} & \multicolumn{1}{c}{$I_i$ (cm$^{-1})$}\\
\hline
$5\mathrm{s}n\mathrm{s}\,^1\mathrm{S}_0$ &
1: $5\mathrm{s}_{1/2}n\mathrm{s}_{1/2}$ & 45932.2002\\ &
2: $4\mathrm{d}_{5/2}n\mathrm{d}_{5/2}$ & 60768.4300\\ &
3: $4\mathrm{d}_{3/2}n\mathrm{d}_{3/2}$ & 60488.0900 \\[1mm]
$5\mathrm{s}n\mathrm{s}\,^3\mathrm{S}_1$ &
1: $5\mathrm{s}n\mathrm{s}\,^3\mathrm{S}_1$ & 45932.2002\\ &
2: $5\mathrm{p}n\mathrm{p}\,^3\mathrm{P}_1$ & 70048.1100\\[1mm]
$5\mathrm{s}n\mathrm{p}\,^1\mathrm{P}_1^{\rm o}$ & 
1: $5\mathrm{s}n\mathrm{p}\,^1\mathrm{P}_1^{\rm o}$& 45932.2002\\ &
2: $4\mathrm{d}n\mathrm{p}\,^1\mathrm{P}_1^{\rm o}$ & 60628.2600\\[1mm]
$5\mathrm{s}n\mathrm{p}\,^3\mathrm{P}_0^{\rm o}$ & 
1: $5\mathrm{s}n\mathrm{p}\,^3\mathrm{P}_0^{\rm o}$& 45932.2002\\ &
2: $4\mathrm{d}n\mathrm{p}\,^3\mathrm{P}_0^{\rm o}$ & 60628.2600\\[1mm]
$5\mathrm{s}n\mathrm{p}\,^3\mathrm{P}_1^{\rm o}$ & 
1: $5\mathrm{s}n\mathrm{p}\,^3\mathrm{P}_1^{\rm o}$& 45932.2002\\ &
2: $4\mathrm{d}n\mathrm{p}\,^3\mathrm{P}_1^{\rm o}$ & 60628.2600\\[1mm]
$5\mathrm{s}n\mathrm{p}\,^3\mathrm{P}_2^{\rm o}$ & 
1: $5\mathrm{s}n\mathrm{p}\,^3\mathrm{P}_2^{\rm o}$& 45932.2002\\ &
2: $4\mathrm{d}n\mathrm{p}\,^3\mathrm{P}_2^{\rm o}$ &60628.2600\\[1mm]
$5\mathrm{s}n\mathrm{d}\,^1\mathrm{D}_2$ and $5\mathrm{s}n\mathrm{d}\,^3\mathrm{D}_2$ &
1: $5\mathrm{s}_{1/2}n\mathrm{d}_{5/2}$ & 45932.2002\\ &
2: $5\mathrm{s}_{1/2}n\mathrm{d}_{3/2}$ & 45932.2002\\ &
3: $4\mathrm{d}_{5/2}n\mathrm{s}_{1/2}$ & 60768.4300\\ &
4: $4\mathrm{d}_{3/2}n\mathrm{s}_{1/2}$ & 60488.0900\\ &
5: $5\mathrm{p}n\mathrm{p}\,^1\mathrm{D}_2$ & 70048.1100 \\ &
6: $4\mathrm{d}n\mathrm{d}\,^3\mathrm{P}_2$ & 60628.2600 \\[1mm]
$5\mathrm{s}n\mathrm{d}\,^3\mathrm{D}_1$ & 
1: $5\mathrm{s}n\mathrm{d}\,^3\mathrm{D}_1$& 45932.2002\\ &
2: $4\mathrm{d}n\mathrm{s}\,^3\mathrm{D}_1$ & 60628.2600\\[1mm]
$5\mathrm{s}n\mathrm{d}\,^3\mathrm{D}_3$ & 
1: $5\mathrm{s}n\mathrm{d}\,^3\mathrm{D}_3$& 45932.2002\\ &
2: $4\mathrm{d}n\mathrm{s}\,^3\mathrm{D}_3$ & 60628.2600\\ &
3: $4\mathrm{d}n\mathrm{d}\,^3\mathrm{D}_3$ & 60628.2600\\[1mm]
$5\mathrm{s}n\mathrm{f}\,^1\mathrm{F}_3^{\rm o}$ & 
1: $5\mathrm{s}n\mathrm{f}\,^1\mathrm{F}_3^{\rm o}$& 45932.2002\\ &
2: $4\mathrm{d}n\mathrm{p}\,^1\mathrm{F}_3^{\rm o}$ & 60628.2600
\end{tabular}
\end{ruledtabular}
\label{table:channels}
\end{table}

\begin{table*}[htbp]
\caption{The experimental data used to construct the present MQDT models.}
\begin{ruledtabular}
\begin{tabular}{ll}
Series & \multicolumn{1}{l}{References}\\
\hline
$5\mathrm{s}n\mathrm{s}\,^1\mathrm{S}_0$ &
$n=7$--$9$: \cite{Sansonetti2010}; $n=10$--$24$, $26$, $28$, $30$ : \cite{Beigang1982a}; $n=25$, $27$, $29$: \cite{Rubbmark1978}. Perturber: \cite{Sansonetti2010}.
\\[0mm]
$5\mathrm{s}n\mathrm{s}\,^3\mathrm{S}_1$ &
(a) $n=7$--$12$: \cite{Sansonetti2010}; $n=13$--$18$: \cite{Beigang1982a}; $n=19$--$23$: \cite{Kunze1993}.\\&
(b) $n=7$--$12$: \cite{Sansonetti2010}; $n=13$--$23$: \cite{Couturier2019}.
\\[0mm]
$5\mathrm{s}n\mathrm{p}\,^1\mathrm{P}_1^{\rm o}$ & 
$n=6$--$20$: \cite{Sansonetti2010}; $n=21$--$29$: \cite{Rubbmark1978}. Perturber: \cite{Sansonetti2010}.
\\[0mm]
$5\mathrm{s}n\mathrm{p}\,^3\mathrm{P}_0^{\rm o}$ & 
$n=6$, $7$: \cite{Sansonetti2010}; $n=8$--$12$, $14$, $15$: \cite{Esherick1977}. Perturber: \cite{Sansonetti2010}.
\\[0mm]
$5\mathrm{s}n\mathrm{p}\,^3\mathrm{P}_1^{\rm o}$ &
$n=6$, $7$: \cite{Sansonetti2010}; $n=8$--$12$, $14$, $15$: \cite{Armstrong1979}. Perturber: \cite{Sansonetti2010}.
\\[0mm]
$5\mathrm{s}n\mathrm{p}\,^3\mathrm{P}_2^{\rm o}$ & 
$n=6$, $7$: \cite{Sansonetti2010}; $n=8$--$12$, $14$, $15$: \cite{Armstrong1979}. Perturber: \cite{Sansonetti2010}.
\\[0mm]
$5\mathrm{s}n\mathrm{d}\,^1\mathrm{D}_2$ and $5\mathrm{s}n\mathrm{d}\,^3\mathrm{D}_2$ &
Singlet states:
$n=8$, $9$: \cite{Sansonetti2010}; $n=10$, $27$: \cite{Esherick1977};
$n=11$--$13$, $17$--$26$
and $27$--$30$: \cite{Beigang1982a}; 
$n=14$--$16$: \cite{Couturier2019}.\\&
Triplet states:
$n=8$, $9$: \cite{Sansonetti2010}; $n=10$, $11$: \cite{Esherick1977};
$n=12$--$30$: \cite{Couturier2019}.
Perturber: \cite{Esherick1977}.
\\
$5\mathrm{s}n\mathrm{d}\,^3\mathrm{D}_1$ &
$n=12$--$15$ and $17$--$50$: \cite{Couturier2019}.
\\[0mm]
$5\mathrm{s}n\mathrm{d}\,^3\mathrm{D}_3$ & 
$n=6$--$8$: \cite{Sansonetti2010}; $n=9$--$21$ and $23$--$29$: \cite{Beigang1982b}.
\\[0mm]
$5\mathrm{s}n\mathrm{f}\,^1\mathrm{F}_3^{\rm o}$ &
$n=4$--$20$: \cite{Sansonetti2010}; $n=21$--$29$: \cite{Rubbmark1978}. Perturber: \cite{Sansonetti2010}.
\end{tabular}
\end{ruledtabular}
\label{table:data}
\end{table*}


\begingroup
\squeezetable
\begin{table}
\caption{The MQDT parameters for the models considered in this work. The parameters not specified in the table were not used as fitting parameters and have a zero value. The numbers between brackets indicate the powers of ten.}
\begin{ruledtabular}
\begin{tabular}{lccrr}
Series & $i$ & $j$ & \multicolumn{1}{c}{$K_{ij}^{(0)}$ and $K_{ji}^{(0)}$} & \multicolumn{1}{c}{$K_{ii}^{(1)}$}\\
\hline
$5\mathrm{s}n\mathrm{s}\,^1\mathrm{S}_0$
& 1 & 1 &   $1.051261[+0]$ & $8.763911[-1]$\\ 
& 1 & 2 &   $3.759864[-1]$ \\ 
& 1 & 3 &  $-2.365485[-2]$ \\ 
& 2 & 2 &  $-6.400925[-1]$ &$4.042584[-1]$\\ 
& 2 & 3 &  $-2.063825[-4]$ \\ 
& 3 & 3 &  $ 3.009087[+0]$ & $-1.722631[+1]$\\[1mm]
$5\mathrm{s}n\mathrm{s}\,^3\mathrm{S}_1$ (a)
& 1 & 1 & $-3.481970[+1]$& $-1.509809[+1]$ \\
& 1 & 2 & $-1.541241[+2]$& \\
& 2 & 2 & $-6.401689[+2]$& $ 3.312806[+2]$ \\[1mm]
$5\mathrm{s}n\mathrm{s}\,^3\mathrm{S}_1$ (b)
& 1 & 1 & $-1.039244[+2]$& $-2.766912[+1]$ \\
& 1 & 2 & $-1.334517[+2]$& \\
& 2 & 2 & $-1.680452[+2]$& $ 5.517184[+1]$
\\[1mm]
$5\mathrm{s}n\mathrm{p}\,^1\mathrm{P}_1^{\rm o}$
& 1 & 1 & $ 1.116809[+1]$& $-9.097862[-1]$ \\
& 1 & 2 & $ 1.616933[+1]$& \\
& 2 & 2 & $ 2.239617[+1]$& $ 4.272626[+0]$
\\[1mm]
$5\mathrm{s}n\mathrm{p}\,^3\mathrm{P}_0^{\rm o}$
& 1 & 1 & $-4.009565[-1]$& $ 1.039923[+0]$ \\
& 1 & 2 & $-2.220569[-1]$& \\
& 2 & 2 & $-4.025180[-1]$& $-1.021696[+0]$
\\[1mm]
$5\mathrm{s}n\mathrm{p}\,^3\mathrm{P}_1^{\rm o}$
& 1 & 1 & $-4.199067[-1]$& $ 1.082615[+0]$ \\
& 1 & 2 & $-2.292304[-1]$& \\
& 2 & 2 & $-3.526179[-1]$& $-1.304779[+0]$
\\[1mm]
$5\mathrm{s}n\mathrm{p}\,^3\mathrm{P}_2^{\rm o}$
& 1 & 1 & $-4.531133[-1]$& $ 1.050866[+0]$ \\
& 1 & 2 & $-2.179619[-1]$& \\
& 2 & 2 & $-5.285102[-1]$& $-4.051199[-1]$
\\[1mm]
$5\mathrm{s}n\mathrm{d}\,^1\mathrm{D}_2$ and 
& 1 & 1 & $-3.853883[-1]$& $-1.775326[+0]$ \\
$5\mathrm{s}n\mathrm{d}\,^3\mathrm{D}_2$
& 1 & 2 & $ 2.308103[-1]$& \\
& 1 & 3 & $-2.996898[-1]$& \\
& 1 & 4 & $ 6.248391[-1]$& \\
& 1 & 5 & $-2.381621[-1]$& \\
& 1 & 6 & $-8.944624[-2]$& \\
& 2 & 2 & $-4.881877[-1]$& $ 2.052554[+0]$ \\
& 2 & 3 & $-6.411698[-1]$& \\
& 2 & 4 & $ 8.101262[-6]$& \\
& 2 & 5 & $-4.849582[-1]$& \\
& 2 & 6 & $ 2.427350[-3]$& \\
& 3 & 3 & $ 1.136225[+0]$& $ 4.733804[+0]$ \\
& 3 & 4 & $ 2.078805[-1]$& \\
& 4 & 4 & $ 1.123831[+0]$& $ 3.989162[+0]$ \\
& 5 & 5 & $ 6.117878[-1]$& $ 5.292869[+0]$ \\
& 6 & 6 & $ 2.205400[+0]$& $ 6.079562[+0]$
\\[1mm]
$5\mathrm{s}n\mathrm{d}\,^3\mathrm{D}_1$
& 1 & 1 & $-7.403359[-1]$& $ 9.684681[-1]$ \\
& 1 & 2 & $ 5.504572[-1]$& \\
& 2 & 2 & $ 1.461400[+0]$& $ 2.777353[-1]$
\\[1mm]
$5\mathrm{s}n\mathrm{d}\,^3\mathrm{D}_3$
& 1 & 1 & $-7.793857[-1]$& $ 1.071997[+0]$ \\
& 1 & 2 & $ 4.360198[-1]$& \\
& 1 & 3 & $ 2.229788[-1]$& \\
& 2 & 2 & $ 1.212314[+0]$& $ 8.514161[+0]$ \\
& 2 & 3 & $-1.683225[-4]$& \\
& 3 & 3 & $-2.238265[-1]$& $ 5.544426[+0]$
\\[1mm]
$5\mathrm{s}n\mathrm{f}\,^1\mathrm{F}_3^{\rm o}$
& 1 & 1 & $ 1.711631[-1]$& $-3.530368[-1]$ \\
& 1 & 2 & $ 4.505951[-1]$& \\
& 2 & 2 & $-6.978294[-1]$& $-1.318505[+0]$\\
\end{tabular}
\end{ruledtabular}
\label{table:results}
\end{table}
\endgroup

Our revised sets of MQDT parameters are presented in Table~\ref{table:results}, quoted to a sufficient number of figures to avoid any need of re-optimizing these parameters before use in future investigations. The corresponding theoretical energies are compared to the experimental energies in the Supplementary Material accompanying this paper. The $\chi^2$ values characterising the quality of the fit and plots of channel fractions can also be found in the Supplementary Material. As in \cite{Vaillant2014}, some of the models considered involve jj-coupled channels; for such cases, the Supplementary Material contains both plots of the jj-coupled channel fractions $|Z_i|^2$ and plots of the corresponding LS-coupled channel fractions $|Z_{\bar{\alpha}}|^2$, the $Z_{\bar{\alpha}}$ coefficients being calculated from the $Z_i$ coefficients by application of the jj to LS recoupling transformation \cite{Lee1973}.

Brief comments on each series are given in the rest of this paper. The Lu-Fano plots presented in \cite{Vaillant2014} are correct as published and are not considered below.


\subsection{$5\mathrm{s}n\mathrm{s}\, ^{1}\mathrm{S}_0$ states}

This series was described by a 3-channel model in \cite{Vaillant2014}, in which the $5\mathrm{s}n\mathrm{s}\, ^{1}\mathrm{S}_0$ channel was complemented by the $4\mathrm{d}_{3/2}n\mathrm{d}_{3/2}$ and $4\mathrm{d}_{5/2}n\mathrm{d}_{5/2}$ jj-coupled channels in view of the importance of the 4d$n$d configuration in these states \cite{Aymar1987,FroeseFischer1981,Aspect1984}. We have re-fitted the data using the same model, now with the updated values of $I_s$ and $\tilde{R}$ mentioned above. We found similar results as in \cite{Vaillant2014}. They are presented in Table~\ref{table:results} and in the Supplementary Material, for completeness. 

The low lying states, in our calculation, have a considerably stronger $4\mathrm{d}_{5/2}n\mathrm{d}_{5/2}$ character than a $4\mathrm{d}_{3/2}n\mathrm{d}_{3/2}$ character, whereas the perturber at $44525.838$~cm$^{-1}$, which is generally identified with the $4\mathrm{d}^2\, ^{3}\mathrm{P}_0$ state \cite{Sansonetti2010,Hassouneh2022}, has a predominant $4\mathrm{d}_{3/2}n\mathrm{d}_{3/2}$ character: for this state, $|Z_i|^2 = 0.82$ for the $4\mathrm{d}_{3/2}n\mathrm{d}_{3/2}$ channel and almost 0 for the $4\mathrm{d}_{5/2}n\mathrm{d}_{5/2}$ channel, giving, in terms of mixing coefficients for the LS-coupled channels,  $|Z_{\bar{\alpha}}|^2 = 0.33$ for the $4\mathrm{d}n\mathrm{d}\, ^{1}\mathrm{S}_0$ channel and 0.49 for the $4\mathrm{d}n\mathrm{d}\, ^{3}\mathrm{P}_0$ channel.

\subsection{{$5\mathrm{s}n\mathrm{s}\, ^{3}\mathrm{S}_1$} states}

We present two sets of results for this series. The first one, set (a), is based on the same 2-channel model and same experimental data as used in \cite{Vaillant2014}. Our calculations confirm the channel fractions obtained in that previous work. However, the MQDT parameters reported in \cite{Vaillant2014} are incorrect.

\begin{figure}[htbp] 
\centering
\includegraphics[width=0.95\columnwidth]{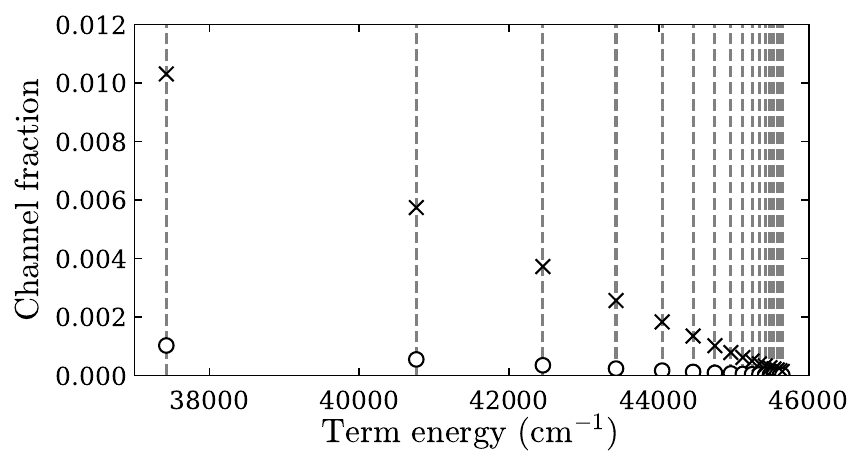}
\caption{Channel fractions for the $5\mbox{s}n\mbox{s}\,^3\mbox{S}_1$ states of Sr. The positions of the relevant experimental bound state energies are indicated by dashed lines. Open circles and crosses: $5\mbox{p}n\mbox{p}\,^3\mbox{P}_1$ channel: the open circles indicate the results obtained with the data set (a), the crosses the results obtained with the data set (b).
}
\label{fig:3S1}
\end{figure}

The second set of results, set (b), is also based on that 2-channel model but uses the recent high precision data of \cite{Couturier2019} for the excited states from $n = 13$ upwards, which have an experimental error of $3 \times 10^{-5}$~cm$^{-1}$. Fitting these recent results proved difficult, though, without increasing their error. We therefore set $\alpha_n^{\rm exp}=0.001$~cm$^{-1}$ for these states, thereby reducing the discrepancy with the error on the energies of the lower lying states (up to 0.35~cm$^{-1}$) for which no high precision results are available, The difficulty may reflect the limitations of the model (e.g., the limited flexibility of the functional form chosen for the energy dependence of the MQDT parameters) or the limitations of the quantum defect method in general (e.g., the assumption that the outer electron interacts with the core through a pure $1/r$ potential, thus neglecting any multipolar interaction \cite{Seaton1983}).
The admixture of the 5p$n$p channel is considerably larger in the results for that model than in the results for model (a) (Fig.~\ref{fig:3S1}). However, this admixture increases monotonically and very regularly as $n$ decreases, which, together with its larger importance, suggests that it is an artefact of the model rather than a physically significant feature. As such, we found no reason to disagree with the conclusion of \cite{Aymar1987} that this series is unperturbed below the first ionisation threshold.


\subsection{$5\mathrm{s}n\mathrm{p} \, ^{1}\mathrm{P}_1^{\rm  o}$ states}

Corrected MQDT parameters for the same model as in \cite{Vaillant2014} are given in Table~\ref{table:results}. The present calculations confirm the channel fractions presented in Fig.~5(a) of that reference.

\subsection{$5\mathrm{s}n\mathrm{p} \, ^{3}\mathrm{P}_0^{\rm  o}$, $^{3}\mathrm{P}_1^{\rm  o}$ and $^{3}\mathrm{P}_2^{\rm  o}$ states}

\begin{figure}[htbp] 
\centering
\includegraphics[width=0.95\columnwidth]{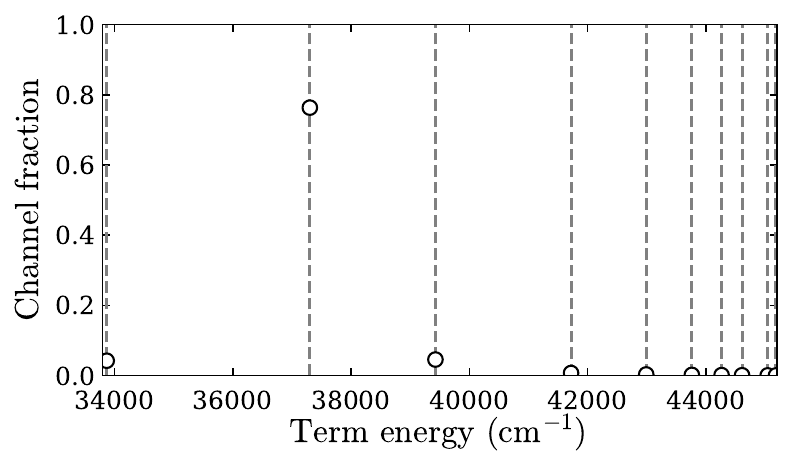}
\caption{Channel fractions for the $5\mbox{s}n\mbox{p}\,^3\mbox{P}_1^{\rm o}$ states of Sr. The positions of the relevant experimental bound state energies are indicated by dashed lines. Open circles: $4\mbox{d}n\mbox{p}\,^3\mbox{P}_1^{\rm o}$ channel.
}
\label{fig:3P1}
\end{figure}

Three-channel models based on the 5s$n$p, 4d$n$p and 4d$n$f configurations were used for analysing the triplet~P series in \cite{Vaillant2014}. However, we found, in the present calculations, that the 4d$n$f channel is not necessary and that two-channel models fit these series satisfactorily within the current experimental errors (we obtained reduced $\chi^2$ values of 2.0, 2.2 and 1.3 for the $J=0$, $J=1$ and $J=2$ series, respectively).  An updated plot of channel fractions superseding Fig.~5(b) of \cite{Vaillant2014} is presented in Fig.~\ref{fig:3P1}. The state with a high admixture of the 4d$n$p channel, at 37302.731~cm$^{-1}$, is the $4\mbox{d}5\mbox{p}\,^3\mbox{P}_1^{\rm o}$ perturber. 

\subsection{{$5\mathrm{s}n\mathrm{d}\, ^{1}\mathrm{D}_2$ and 
$^{3}\mathrm{D}_2$ states}}

High precision results have recently become available for some of the the $^{1}\mathrm{D}_2$ states and many of the $^{3}\mathrm{D}_2$ states \cite{Couturier2019}. Using those, with the experimental errors magnified as mentioned above, lead to the MQDT parameters listed in Table~\ref{table:results} for the same 6-channel model as 
in \cite{Vaillant2014}. The corresponding channel fractions are presented in the Supplementary Material. Our results are similar to those shown in Fig.~7 of \cite{Vaillant2014}, with only minor differences for the
$5\mathrm{p}n\mathrm{p}\, ^{1}\mathrm{D}_2$ and 
$4\mathrm{d}n\mathrm{d}\, ^{3}\mathrm{P}_2$ channels. They confirm this previous analysis.

\subsection{$5\mathrm{s}n\mathrm{d}\, ^{3}\mathrm{D}_1$ states}


A 3-channel model was used in \cite{Vaillant2014} for this series. However, we found that a 2-channel model is sufficient for fitting the data now available \cite{Couturier2019}. The MQDT parameters listed in Table~\ref{table:results} yield a reduced $\chi^2$ of 0.75 when the errors on the experimental energies are enlarged as mentioned above. The resulting channel fractions are very similar to those shown in Ffg.~9(a) of \cite{Vaillant2014} for the 4d$n$s channel, although the model is different. 

Like previous investigators \cite{Beigang1983}, we found that the 5s16d state had to be excluded from the data set in order to obtain a satisfactory fit. We note, in this respect, that the two measurements of the energy of that state gave results in good agreement with each other (45341.36(15)~cm$^{-1}$ \cite{Beigang1982b} and 45341.24788(5)~cm$^{-1}$ \cite{Couturier2019}). Perturbation of this series by an experimentally unobserved 4d6s state has been found to be significant in {\it ab initio} calculations \cite{Aymar1987}. On this basis, we assign the perturbing channel included in our calculation to the 4d$n$s channel rather than the 4d$n$d channel. The channel fractions found in \cite{Vaillant2014} and in the present work for this 4d$n$s channel peak in the vicinity of the 5s16d state, which suggests that the poor fit of that particular state arises from its perturbation by the 4d6s state. A similar situation is also found in the $5\mathrm{s}n\mathrm{d}\, ^{3}\mathrm{D}_3$ series, as discussed below.

\subsection{$5\mathrm{s}n\mathrm{d}\, ^{3}\mathrm{D}_3$ states}

No new experimental results have been published for the $5\mathrm{s}n\mathrm{d}\, ^{3}\mathrm{D}_3$ series since its study in \cite{Vaillant2014}. We therefore used the same data and the same 3-channel model for these states as in that previous work --- i.e., the data from \cite{Sansonetti2010} and \cite{Beigang1982b}. As observed in \cite{Beigang1983} and in our previous work, the model predicts a value of the 5s22d energy particularly discrepant with the experimental energy; we have therefore excluded this state from the fit. 

\begin{figure}[htbp]
\centering
\includegraphics[width=0.95\columnwidth]{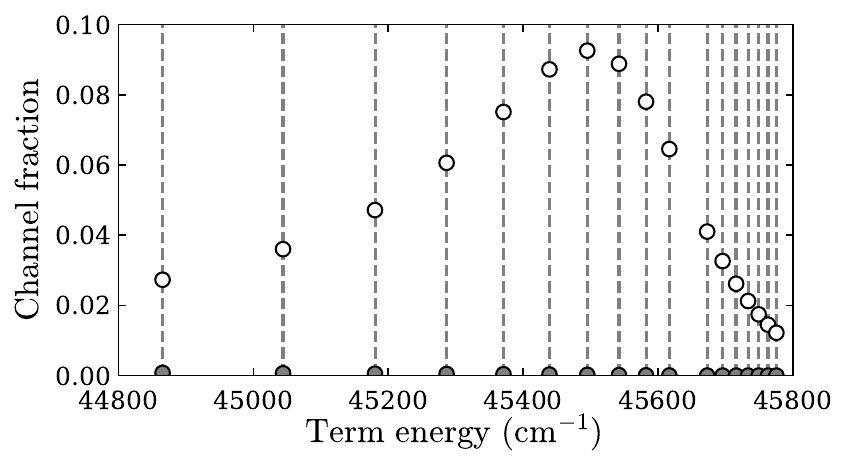}
\caption{Channel fractions for the $5\mbox{s}n\mbox{d}\,^3\mbox{D}_3$ states of Sr. The positions of the relevant experimental bound state energies are indicated by dashed lines. Open circles: $4\mbox{d}n\mbox{s}\,^3\mbox{D}_3$ channel. Filled circles: $4\mbox{d}n\mbox{d}\,^3\mbox{D}_3$ channel.
}
\label{fig:3D3}
\end{figure}

The resulting channel fractions differ from those found in \cite{Vaillant2014}. They are presented in Fig.~\ref{fig:3D3}, which supersedes Ffg.~9(b) of \cite{Vaillant2014}. Similarly to the case of the $J=1$ states, one of the two perturbing channels is significant over the entire set of states, here peaking in the vicinity of the 5s22d state. This feature and the difficulties of fitting that state may also be due to a perturbation by the unobserved 4d6s state. As in \cite{Vaillant2014}, and owing to the importance of that perturber, we tentatively identify the channel with the largest fraction as the 4d$n$s channel (since they have the same ionisation limit, the two channels converging to the 4d threshold cannot be identified unambiguously).   

\subsection{{$5\mathrm{s}n\mathrm{f}\, ^{1}\mathrm{F}_3^{\rm  o}$  states}}

\begin{figure}[htbp]
\centering
\includegraphics[width=0.95\columnwidth]{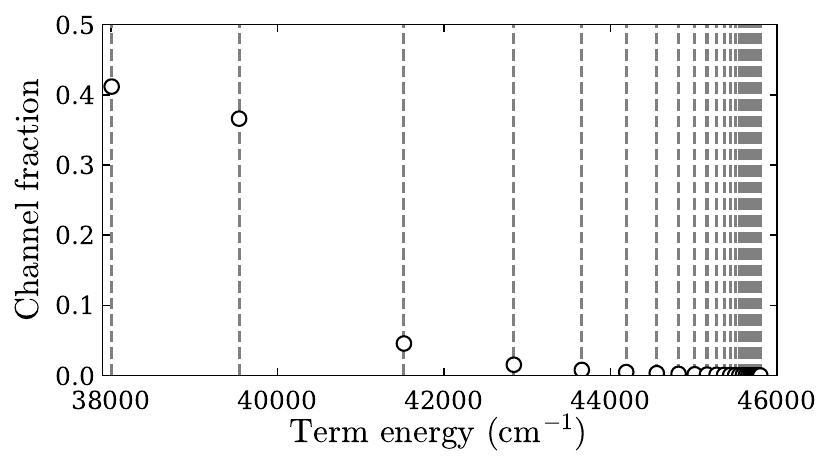}
\caption{Channel fractions for the $5\mbox{s}n\mbox{f}\,^1\mbox{F}_3^{\rm o}$ states of Sr. The positions of the relevant experimental bound state energies are indicated by dashed lines. Open circles: $4\mbox{d}n\mbox{p}\,^1\mbox{F}_3^{\rm o}$ channel.
}
\label{fig:1F3}
\end{figure}

A corrected set of MQDT parameters is given in Table~\ref{table:results} for the same model as in \cite{Vaillant2014}. The resulting channel fractions differ significantly from those obtained in this previous work. They are presented in Fig.~\ref{fig:1F3}. They are in agreement with previous investigations in regard to the importance of the 4d5p perturber and its admixture in the states conventionally labelled $4\mathrm{d}5\mathrm{p}\, ^{1}\mathrm{F}_3^{\rm  o}$, at 38007.742~cm$^{-1}$, and $5\mathrm{s}4\mathrm{f}\, ^{1}\mathrm{F}_3^{\rm  o}$, at 39539.013~cm$^{-1}$ \cite{Hansen1977,Vaeck1988,Kompitsas1990}.


\acknowledgements
We are grateful to Patrick Cheinet, Jonas Cieslik and Rick van Bijnen for having alerted us to the existence of errors in our original paper, to Thomas Killian for having communicated the results of \cite{Brienza2023} in advance of their publication, and to Andreas Kruckenhauser for very useful advice on the calculation of the mixing coefficients for the D$_2$ series. 

\bibliographystyle{iopart-num}
\bibliography{mqdt}

\providecommand{\newblock}{}
\begin{thebibliography}{10}
\expandafter\ifx\csname url\endcsname\relax
  \def\url#1{{\tt #1}}\fi
\expandafter\ifx\csname urlprefix\endcsname\relax\def\urlprefix{URL }\fi
\providecommand{\eprint}[2][]{\url{#2}}

\bibitem{Vaillant2014}
Vaillant C~L, Jones M~P~A and Potvliege R~M 2014 {\em J. Phys. B: At. Mol. Opt.
  Phys.\/} {\bf 47} 155001

\bibitem{Vaillant2015}
Vaillant C~L, Potvliege R~M and Jones M~P~A 2015 {\em Phys. Rev. A\/} {\bf 92}
  042705

\bibitem{Couturier2019}
Couturier L, Nosske I, Hu F, Tan C, Qiao C, Jiang Y~H, Chen P and
  Weidem{\"u}ller M 2019 {\em Phys. Rev. A\/} {\bf 99} 022503

\bibitem{Tan2022}
Tan C, Hu F, Jiang Y, Weidem{\"u}ller M and Zhu B 2022 {\em Chinese. Phys.
  Lett.\/} {\bf 39} 093202

\bibitem{Brienza2023}
Brienza R~A, Lu Y, Wang C, Kanungo S~K, Killian T~C, Dunning F~B,
  Burgd{\"o}rfer J and Yoshida S 2023 {\em Phys. Rev. A\/} {\bf 108} 022815

\bibitem{Beigang1982a}
Beigang R, L\"ucke K, Timmermann A and West P~J 1982 {\em Opt. Comm.\/} {\bf
  42} 19

\bibitem{Seaton1983}
Seaton M~J 1983 {\em Rep. Prog. Phys.\/} {\bf 46} 167

\bibitem{PotvliegeCPC2023}
Potvliege R~M 2024 {\em Comput. Phys. Commun.\/} {\bf 300} 109172

\bibitem{mqdtfitGitHub}
\url{ https://github.com/durham-qlm/mqdtfit}

\bibitem{Sansonetti2010}
Sansonetti J~E and Nave G 2010 {\em J. Phys. Chem. Ref. Data\/} {\bf 39} 033103

\bibitem{Rubbmark1978}
Rubbmark J~R and Borgstr{\"o}m S~A 1978 {\em Phys. Scripta\/} {\bf 18} 196

\bibitem{Kunze1993}
Kunze S, Hohmann R, Kluge H~J, Lantzsch J, Monz L, Stenner J, Stratmann K,
  Wendth K and Zimmer K 1993 {\em Z. Phys. D\/} {\bf 27} 111

\bibitem{Esherick1977}
Esherick P 1977 {\em Phys. Rev. A\/} {\bf 15} 1920

\bibitem{Armstrong1979}
Armstrong J~A, Wynne J~J and Esherick P 1979 {\em J. Opt. Soc. Am.\/} {\bf 69}
  211

\bibitem{Beigang1982b}
Beigang R, L\"ucke K, Schmidt D, Timmermann A and West P~J 1982 {\em Phys.
  Scripta\/} {\bf 26} 183

\bibitem{Lee1973}
Lee C~M and Lu K~T 1973 {\em Phys. Rev. A\/} {\bf 8} 1241

\bibitem{Aymar1987}
Aymar M, Luc-Koenig E and Watanabe S 1987 {\em J. Phys. B: At. Mol. Opt.
  Phys.\/} {\bf 20} 4325

\bibitem{FroeseFischer1981}
Froese~Fischer C and Hansen J~E 1981 {\em Phys. Rev. A\/} {\bf 24} 631

\bibitem{Aspect1984}
Aspect A, Bauche J, Fonseca A~L~A, Grangier P and Roger G 1984 {\em J. Phys. B:
  At. Mol. Phys.\/} {\bf 17} 1761

\bibitem{Hassouneh2022}
Hassouneh O and Salah W 2022 {\em Eur. Phys. J. Plus\/} {\bf 137} 944

\bibitem{Beigang1983}
Beigang R and Schmidt D 1983 {\em Phys. Scripta\/} {\bf 27} 172

\bibitem{Hansen1977}
Hansen J~E and Persson W 1977 {\em J. Phys. B: At. Mol. Opt. Phys.\/} {\bf 10}
  L363

\bibitem{Vaeck1988}
Vaeck N, Godefroid M and Hansen J~E 1988 {\em Phys. Rev. A\/} {\bf 38} 2830

\bibitem{Kompitsas1990}
Kompitsas M, Cohen S, Nicolaides C~A, Robaux O, Aymar M and Camus P 1990 {\em
  J. Phys. B: At. Mol. Opt. Phys.\/} {\bf 23} 2247

\end{thebibliography}
%
%
%
%
%


\end{document}